\documentclass[%
 aps,prl,
 amsmath,amssymb,
 superscriptaddress,
preprint,
preprintnumbers,
showkeys
]{revtex4-2}

\usepackage[english]{babel}

\usepackage{amsmath,amssymb,amsfonts,amsthm}

\usepackage{graphicx}
\usepackage{dcolumn}
\usepackage{bm,color}
\usepackage{epstopdf}
\usepackage{nicefrac}
\usepackage{siunitx}

\begin{document}

\begin{abstract}
	In this Letter, we present investigations of thermal transport in (LaMnO$_3$)$_n$/(SrMnO$_3$)$_n$ superlattices (SLs) with SL periods $\Lambda=2n$ between $2$ and $12$ unit cells. The analysis of the experimental data clearly reveals a minimum in the thermal conductivity $\kappa$ at $\Lambda=6$ u.c. Furthermore, we theoretically estimate the phonon mean free path to be about $3\Lambda$ in our SLs.	These findings clearly show that, in a wide temperature range of $\pm 100\,$K around room temperature, thermal transport in our manganite SLs is coherent. In addition, we demonstrate that antiferromagnetic magnetic ordering can control the degree of coherency. This opens the challenging perspective for active control of coherent transport in correlated matter.
\end{abstract}

\title[]{Coherent phonon transport and minimum of thermal conductivity in LaMnO$_3$/SrMnO$_3$ superlattices}

\date{\today}

\author{Dennis Meyer}
\email{dmeyer@uni-goettingen.de}
\affiliation{I. Physical Institute, Georg-August University of G\"ottingen, Friedrich-Hund-Platz 1, 37077 G\"ottingen, Germany}
\author{Daniel Metternich}
\affiliation{I. Physical Institute, Georg-August University of G\"ottingen, Friedrich-Hund-Platz 1, 37077 G\"ottingen, Germany}
\affiliation{Helmholtz-Zentrum Berlin f\"ur Materialien und Energie,
	Hahn-Meitner-Platz 1,
	14109 Berlin, Germany}
	\author{Pia Henning}
\affiliation{I. Physical Institute, Georg-August University of G\"ottingen, Friedrich-Hund-Platz 1, 37077 G\"ottingen, Germany}
\author{Jan Philipp Bange}
\affiliation{I. Physical Institute, Georg-August University of G\"ottingen, Friedrich-Hund-Platz 1, 37077 G\"ottingen, Germany}
\author{Robert Gruhl}
\affiliation{I. Physical Institute, Georg-August University of G\"ottingen, Friedrich-Hund-Platz 1, 37077 G\"ottingen, Germany}
\affiliation{Experimentalphysik VI, University of Augsburg, Universit\"atsstraße 1, 86159 Augsburg, Germany}
\author{Vitaly Bruchmann-Bamberg}
\author{Vasily Moshnyaga}
\author{Henning Ulrichs}
\affiliation{I. Physical Institute, Georg-August University of G\"ottingen, Friedrich-Hund-Platz 1, 37077 G\"ottingen, Germany}
\email{hulrich@gwdg.de}


\maketitle

\section{\label{intro}Introduction}

Coherent wave propagation gives rise to fascinating transport physics in solids. For instance, Bloch-oscillations \cite{Bloch1929,PhysRevB.46.7252}, and the quantum Hall effect \cite{Ando1975,Wakabayashi1978} rely on long temporal and large spatial coherence of electronic waves. Also bosonic quasi-particles show interesting dynamical phenomena based on coherence. For instance, magnon Bose-Einstein-condensation has been observed in yttrium-iron-garnet (YIG) at room temperature \cite{Demokritov2006}. This macroscopic quantum state is enabled due to the exceptionally long lifetimes of magnons in YIG \cite{CHEREPANOV199381}. A measureable consequence of phonon coherence is the formation of caustic focusing patterns \cite{PhysRevLett.12.641,PhysRevLett.23.416,PhysRevLett.43.1424,Veres_2012}, which are also known for magnons  \cite{PhysRevB.61.11576,PhysRevB.74.214401,Gieniusz2013}. In the absence of scattering processes not conserving the phonon momentum, one can observe the phenomenon of second sound, which is a coherently propagating density variation in a thermally excited phonon gas \cite{PhysRev.131.2013,PhysRevLett.28.1461}. Note that, second sound was also recently reported in a magnon BEC \cite{Bozhko2019,Tiberkevich2019}.

The majority of the examples and phenomena cited above are rather of academic interest, than of great practical importance, since they typically manifest only at cryogenic temperatures or in exotic materials.
The most common physical process related to phonons is thermal transport. At room temperature, it is usually assumed to be diffusive, and well-described by the Fourier law. This picture is certainly appropriate when the spatial scales involved in the heat flow are much larger than the phonon mean free path (mfp) $\xi$. When further downscaling nanoscale systems like the upcoming sub-5nm transistor chips\cite{H.Lee.2006, Cahill.2003, Cahill.2014},  
the wave properties of phonons need be taken into account. Despite this technological relevance, only few clear experimental reports of coherent (of ballistic) phonon thermal tranport at room temperature exist so far \cite{Siemens2010,Luckyanova2012,Ravichandran2013}. In many works, one-dimensional superlattices (SLs) are considered, which are indeed a paradigmatic model system for this specific research topic \cite{PhysRevB.61.3091,Simkin.2000, Chen.2005, Luckyanova2012,Hu2012,Ravichandran2013,PhysRevB.93.045311,Chakraborty2017,PhysRevB.97.085306,Chakraborty_2020}. In their theoretical work \cite{Simkin.2000}, Simkin and Mahan have shown that, when varying the SL period $\Lambda$, coherence should lead to a minimum in the dependence of the thermal conductivity $\kappa$ on $\Lambda$, when $\xi\gg\Lambda$. To our knowledge, this feature has so far only been shown by Ravichandran et al. \cite{Ravichandran2013} in titanate SLs consisting of CaTiO$_3$, BaTiO$_3$ and SrTiO$_3$ and by Saha et al. \cite{Saha.2017} in TiN/(Al,Sc)N metal/semiconductor superlattices.

\par\medskip
Here, we present thermal conductivity measurements on (LaMnO$_3$)$_n$/(SrMnO$_3$)$_n$ superlattices with different atomic layer thicknesses $n=1$ to $6$. In this manganite SL system charge transfer gives rise to interfacial ferromagnetism, whose discovery attracted scientific curiosity in recent years \cite{Mercey1999,Salvador1999,Adamo2008,PhysRevB.77.174409,PhysRevB.78.201102,PhysRevB.80.140405,PhysRevB.81.014410,PhysRevB.86.174427,TANG20138,Keunecke2020}. 
Very recently \cite{meyer2021atomic} we have show that such SLs display a small interfacial phonon scattering, likely enabling coherent thermal transport with $\xi\approx 3\Lambda$. As mentioned above, the smoking gun of coherent transport is the minimum in  $\kappa(\Lambda)$. For this Letter, we have further optimized the sample growth, and our optical setup for measuring thermal transport properties. Going down to the atomic layer limit (SLs with $n=1$), we can now report exactly this feature of a minimum in $\kappa(\Lambda)$. In addition to experimental evidence, and theoretical support for coherent thermal transport, we also show how interaction with antiferromagnetic spin ordering can give rise to a significant reduction of the phonon mfp.


\medskip

\begin{figure*}[ht!]
\includegraphics[width=16cm]{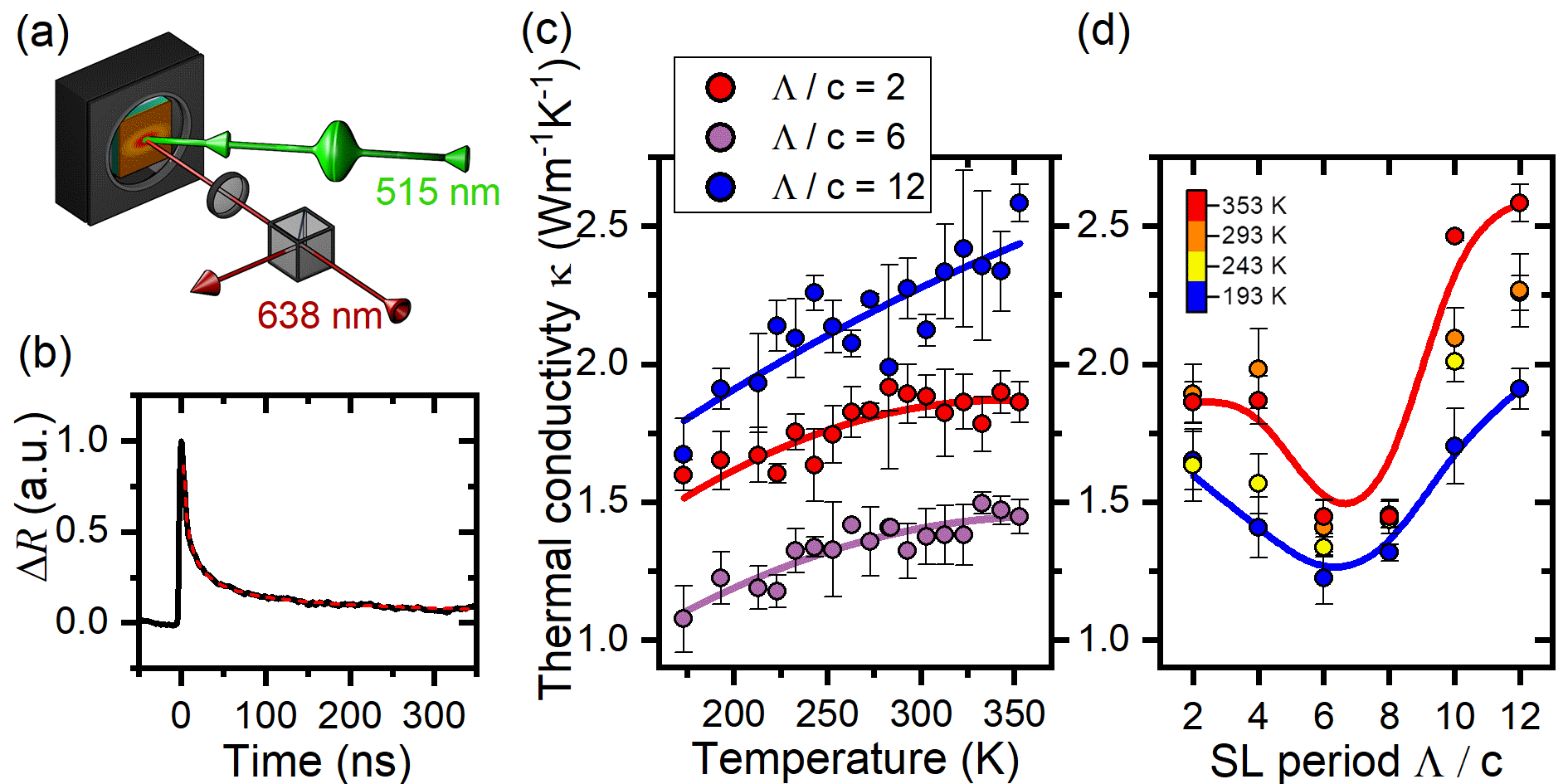}
\caption{Thermal transport in LMO$_n$/SMO$n$ SLs with $n=1$ to $6$. (a) Sketch of the TTR setup. (b) Exemplary TTR raw data obtained at $T=303\,$K from the SL with $\Lambda=6c$ (black line). The dashed red line is a fit of our thermal model \cite{Balageas1986, meyer2021atomic} to the data. (c) Temperature dependence of the thermal conductivity $\kappa$ for SLs with different SL period $\nicefrac{\Lambda}{c} = 2, 6$ and $12$. Solid lines in the background as a guide to the eye. (d) Dependence of $\kappa$ on the SL period $\Lambda$ for selected temperatures as indicated. Solid lines as a guide to the eye. \label{fig:1}}

\end{figure*}
The (LaMnO$_3$)$_n$/(SrMnO$_3$)$_n$ SLs (LMO$_n$SMO$_n$ SLs) have been grown by metalorganic aerosol deposition (MAD)\cite{Jungbauer.2014, Keunecke2020} on (100)-oriented SrTiO$_3$ (STO) substrates. Within this method, in-situ ellipsometry enables atomic-layer resolved growth control. As we have show in detail in \cite{meyer2021atomic}, SLs with $n<10$ display in both LMO, and SMO layers, a pseudo-cubic structure with in-plane lattice constants $a=b=0.39\,$nm adapted from the STO substrate, and out-of-plane lattice constant $c=0.38\,$nm. Most importantly, the oxygen octrahedra tilt mismatch in LMO (as well as in SMO) is absent in SLs with $n<10$. The latter is present in LMO layers inside LMO$_{2n}$SMO$_n$, and is characteristic for bulk LMO. 

\medskip
Thermal transport measurements have been carried out with nanosecond transient thermoreflectance (TTR) \cite{Kaeding1994}.
In our TTR setup \cite{meyer2021atomic} (see Fig. \ref{fig:1}(a) for a sketch), a pulsed nanosecond laser (wave length \SI{515}{\nano\meter}, repetition rate \SI{2}{\kilo\hertz}) is used to heat a \SI{50}{\nano\meter} thick copper film, which was grown on all samples by electron beam evaporation in UHV conditions. Using a confocal beam path, a cw diode laser (\SI{639}{\nano\meter}) is reflected at the sample surface, subsequently converted on a symmetric photodetector into a voltage signal, which is then recorded in real time by an oscilloscope. Since the reflectivity $R\propto T$, this signal includes information about transport of thermal energy from the Cu film, through the SLs, and into the substrate. By means of a custom-build liquid-nitrogen cryostat, we can control the sample temperature between $100\,$K and $400\,$K. Figure \ref{fig:1}(b) shows a typical TTR curve, obtained from the SL with $n=2$ at a  temperature of $T=303\,$K. This curve is an average over $2^{16}$ pump events. After the initial sudden increase of the surface temperature caused by the absorption of the pump pulse, one can see a monotoneous cooling down to the base temperature of the cryostat. To analyze the dynamics quantitatively, we fit a thermal three-layer model to these data \cite{Balageas1986, meyer2021atomic}. From this, we obtain the thermal conductivity $\kappa$ of the SLs.

The central experimental results discussed in this Letter are presented in Fig. \ref{fig:1}(c), where we show the temperature dependence of $\kappa$ of all SLs. Note that, we have determined $\kappa$ as an average from three independent TTR measurements at each temperature from each SL sample. The error bar is the corresponding mean standard deviation. One can generally see an increase of $\kappa$ with increasing temperature, which is typical for manganites \cite{Cohn.1997}. Here, we emphasize that, the SL with the smallest thermal conductivity is not the SL with smallest SL period. Indeed, the SL sample with $\Lambda=6c$ has the minimal thermal conductivity. This important finding is further emphasized in Fig. \ref{fig:1}(d). There, we show $\kappa$ as a function of $\Lambda$ for selected temperatures as indicated. Between $193$ K and $353$ K, we did not detect a systematic shift of the minimum to SL periods larger or smaller than $\Lambda=6c$.

\begin{figure*}[ht!]
\includegraphics[width=16cm]{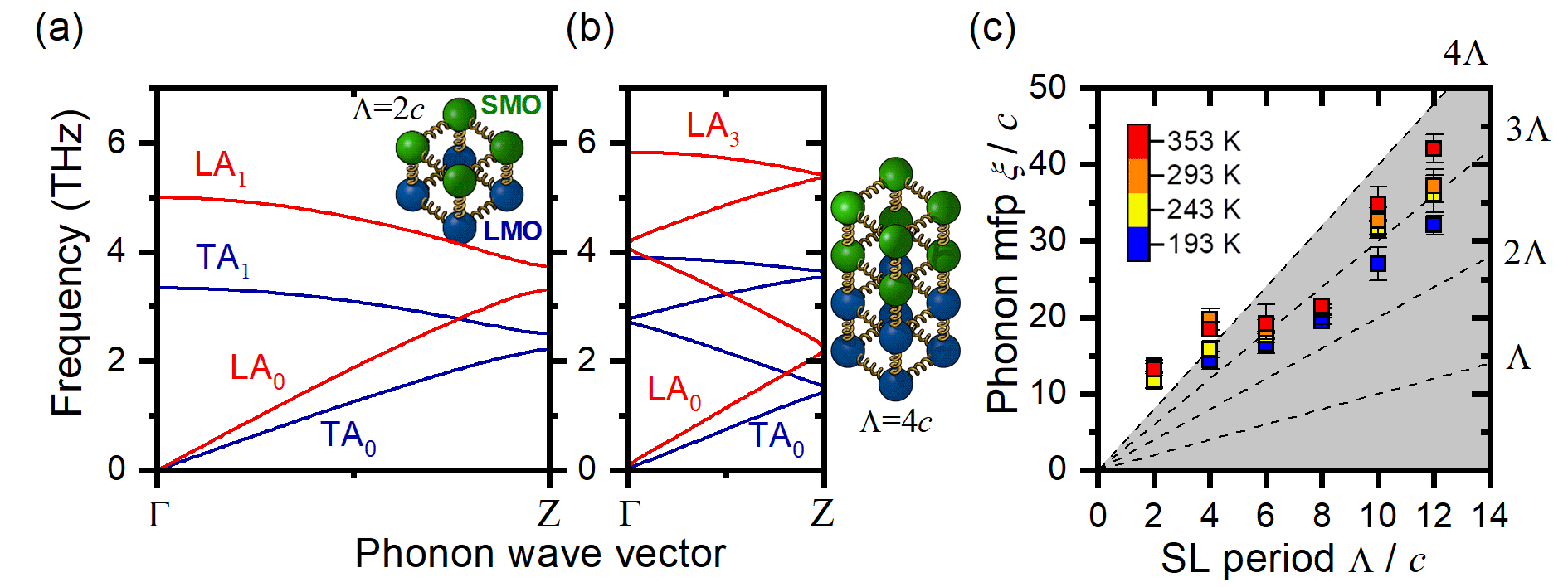}
\caption{Lattice dynamical modelling of thermal transport in SLs. (a) Acoustic phonon spectrum of the SL with $\Lambda=2c$ from $\Gamma=(0,0,0)$ to $\mathrm{Z}=(0,0,\nicefrac{\pi}{\Lambda})$. The symbols (LA$_{j}$, TA$_{j}$) denote the origin of the back-folded bands. The inset depicts a sketch of the BKM, showing effective SMO (green) and LMO (blue) atoms in $2\times2$ unit cells connected by springs, representing harmonic interaction potentials. (b) Acoustic phonon spectrum of the SL with $\Lambda=4c$, again, together with a sketch of the BKM. (c) Phonon mfp as a function of the SL period at different temperatures as indicated. Dashed lines mark different levels of coherence as indicated.} \label{fig:2}
\end{figure*}

To further corroborate this experimental prove of coherent thermal transport, we have applied a Born-von-Kármán-like lattice dynamical model (BKM) \cite{born1914raumgittertheorie}. This model enables us to calculate the acoustic part of the SL phonon spectrum $\omega_i(\mathbf k)$, with $i=1$ to $6n$, and to estimate the phonon mfp $\xi$. Within the BKM, each LMO and SMO unit cell is described as a single effective atom, coupled by harmonic springs with neighbouring effective atoms (see Fig. \ref{fig:2}(a), and (b) for a sketch). More details on the BKM can be found in \cite{meyer2021atomic}. The exemplary spectra in Fig. \ref{fig:2}(a), and (b) refer to SLs with $\Lambda=2c$, and $\Lambda=4c$. In these diagrams, we have traced out a path in reciprocal space from $\Gamma=(0,0,0)$ to the edge of the first SL Brillouin zone (BZ) at $\mathrm{Z}=(0,0,\nicefrac{\pi}{\Lambda})$. In both examples, one can clearly see the back-folding of the phonon bands, and the formation of (small) band gaps at $Z$, and at $\Gamma$. The origin of the bands in the longitudinal or transverse acoustic phonon branches, shifted by a reciprocal lattice vector $\mathbf G_{00j}=(0,0,\nicefrac{j\pi}{\Lambda})$, is denoted by the symbols LA$_j$, and TA$_{j}$. To calculate the phonon thermal conductivity, we evaluate 

\begin{eqnarray}
 \kappa =\sum_i \int_\mathrm{1.BZ}\frac{d^3k}{(2\pi)^3}\hbar\omega \left\vert v_z(\mathbf k)\right\vert \xi \frac{\partial n_\mathrm{BE}}{\partial T},\label{Eq:SLkappa}
\end{eqnarray}

where the summation extends over all phonon branches $i$, $v_z=\frac{\partial\omega_i(\mathbf k)}{\partial k_z}$, and $n_\mathrm{BE}$ is the Bose-Einstein distribution. Following Simkin and Mahan \cite{Simkin.2000}, the phonon mfp $\xi$ enters directly into the integrand in Eq. \eqref{Eq:SLkappa}, and indirectly by assuming complex wave numbers with imaginary part $\xi^{-1}$ for the calculation of the phonon spectrum $\omega_i(\mathbf k)$. Fitting the theoretical expectation according to Eq.\eqref{Eq:SLkappa} to the experimentally determined thermal conductivities, enables us to determine $\xi$ as a function of the SL period $\Lambda$, as shown in Fig. \ref{fig:2}(c). At all temperatures we find a monotonous increase of $\xi$ with increasing $\Lambda$. Moreover, for all SLs $\xi\gg\Lambda$. Note that, $\xi=\Lambda$ can be considered as a boundary, below which thermal transport in a SL is diffusive, and above which it can be called coherent. Thus, the experimental data in conjunction with the lattice dynamical model clearly show the coherent nature of thermal transport in our atomic-scale manganite SLs.

\begin{figure*}[ht!]
\includegraphics[width=16cm]{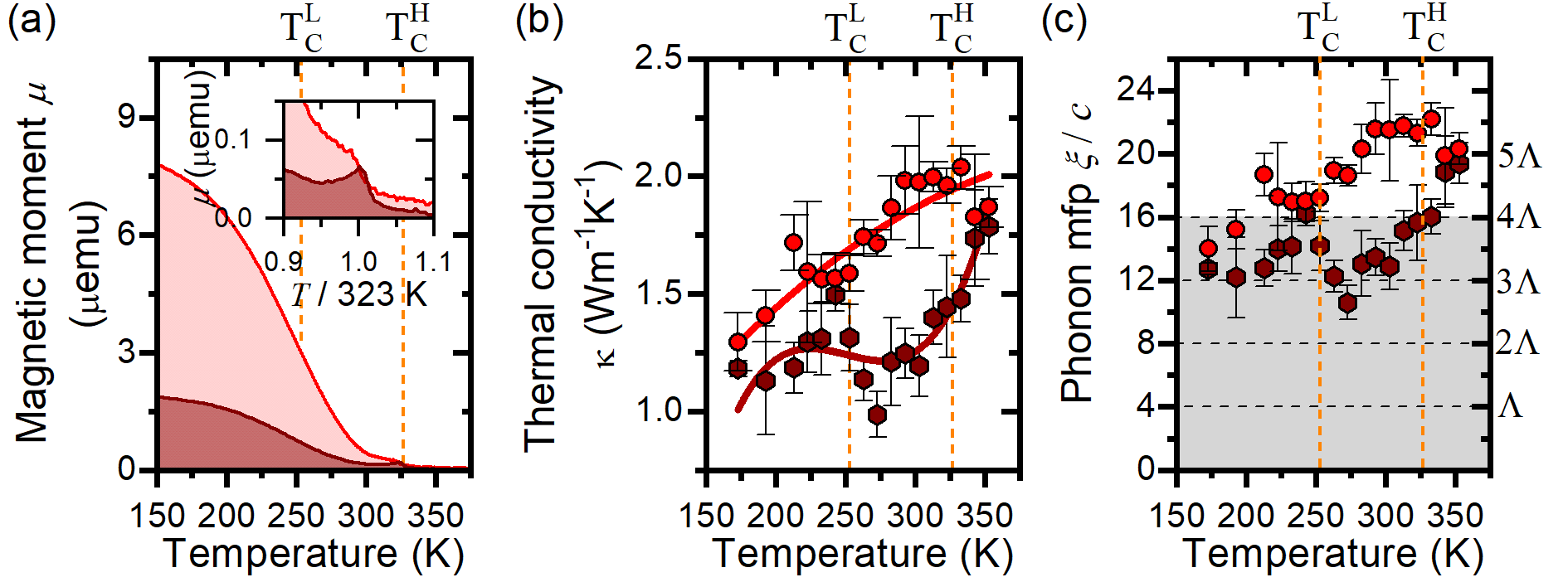}
\caption{Influence of magnetic order on thermal transport. (a) Field-cooled magnetization curves for two nominally identical SLs with $n=2$, measured at a magnetic field of $10\,$mT . The dark red curve refers to the SL with slightly detuned stoichiometry (SL$_4^\delta$), the light red to an SL with correct stoichiometry (SL$_4$). (b) Thermal conductivity as a function of temperature for SL$_4^\delta$ (hexagonal dot) and SL$_4$ (round dot). Solid lines in the background as a guide to the eye. (c) Temperature dependence of the phonon $mfp$ for SL$_4^\delta$ (hexagonal symbol), and for SL$_4$ (circular dot). The greyed out background marks different levels of coherence, analogously to Fig. \ref{fig:2}(c). \label{fig:3}}
\end{figure*}

\medskip
We now show that, by slightly detuning the stoichiometry of a SL with $\Lambda=4c$ (SL$_4^\delta$, increased Sr content, and decreased La content), we can significantly impact the thermal transport via magnetic ordering. In Fig. \ref{fig:3}(a) SQUID magnetometry in terms of field cooled magnetization curves for an ideal SL (SL$_4$), and the detuned SL are shown. The here shown full magnetic moment $\mu(T)$ of 
SL$_4$ features a small kink at about $323(2)\,$K (maximum in the logarithmic derivative $\frac{\partial \mathrm{ln}(\mu)}{\partial T}$), besides the dominant ferromagnetic-to-paramagnetic PT with $T^\text{L}_\mathrm{C}=251(5)\,$K. Following Keunecke et al. \cite{Keunecke2020}, we associate the kink with a   ferromagnetic-to-paramagnetic PT of a high-T FM phase at  $T^\text{H}_{\mathrm{C}}$ and an antiferromagnetic (AFM) phase located at each second LMO/SMO interface between the FM phases. For SL$_4^\delta$, the AFM phase is much more pronounced, causing an overall reduction of $\mu(T)$ compared to  SL$_4$, and a clearly visible local maximum at $323\,$K ($T^\text{H}_\mathrm{C}=327(5)\,$K) [see Fig. \ref{fig:3}(a)]. The TTR measurements on this sample reveal a significantly reduced thermal conductivity, compared to the ideal SL [compare hexagonal dots with red curve in Fig. \ref{fig:3}(b)]. This translates according to our lattice dynamical model into a reduction of the phonon mfp by almost a factor of two closely below $T^\text{H}_{\mathrm{C}}$ [see Fig. \ref{fig:3}(c)]. In the absence of any magnetic ordering at $T\gg T^\text{H}_{\mathrm{C}}$, and when the odering is fully established at $T\ll T^\text{L}_\mathrm{C}$, thermal conductivities of both SLs are approximately equal. Several reasons can relate the reduction of the mfp to the presence of an AFM phase. i) Phonons are likely scattered from AFM domain walls.  ii) Antiferromagnetic magnons may hybridize with acoustic phonons, which opens a new magnetic relaxation pathway, and thereby decreases the mfp \cite{PhysRevLett.125.217201}. Indeed, the observation of an increase in ultrasonic attenuation, explained by i), and ii), was historically one of the first experimental methods to access AFM phase transitions \cite{PhysRev.173.542,PhysRevB.1.3083}.

\medskip

In this Letter, we have presented experimental and theoretical evidence for coherent thermal transport in atomic-scale manganite superlattices. We here emphasize that, in difference to the titanate SLs considered by Ravichandran et al. \cite{Ravichandran2013}, transport physics in manganites is generally strongly influenced by magnetic degrees of freedom \cite{PhysRevLett.71.2331,Jin413, Cohn.1997}. In our SLs, an enhanced spin lattice interplay manifests in a reduced thermal conductivity when increasing the spatial extend of the AFM phase. At the AFM to PM phase transition a passive control of the thermal transport manifests in the data. For future devices implementing actively controlled elements for thermal circuitry, like for example a thermal switch or diode \cite{Peyrard2006}, nano-structured correlated manganites offer an interesting perspective. Taking a detour through the spin system, they may allow to externally control coherency in thermal transport by magnetic fields or spin currents.

\medskip

\begin{acknowledgments}
This research was funded by the Deutsche Forschungsgemeinschaft (DFG, German Research Foundation) - 217133147/SFB 1073, project A02.
\end{acknowledgments}




\setcounter{secnumdepth}{2}

\bibliography{sources,citavi}
\end{document}